# Deuteronomy 2.0: Record Caching and Latch Freedom

David Lomet
Redmond, WA, USA
dlomet@msn.com

## ABSTRACT

The Deuteronomy transactional key-value store is unique architecturally in providing separation between transaction functionality-its Transactional Component (TC) and data management- its Data Component (DC). It is unique in technology by (1) supporting record caching, a smaller unit than the traditional page; and (2) protecting resources during concurrent execution using a latch-free approach. Both technologies are enabled by delta updating. This paper explains how record caching improves cache cost/performance. It also shows how a new latch-free approach makes implementation easier and improves performance.

## 1   Introduction

Deuteronomy has many implementation features that are unique to key-value store architecture [2] and algorithmic design [3,4,5]. We focus here on how delta updating enables two aspects of Deuteronomy and discuss both why they lead to high performance and how to make them even better. These aspects represent a new way of thinking about and implementing key-value stores. Delta updating enables updates without the need to make changes in-place. This pervades Deuteronomy, enabling record caching and incremental page updating in a latch-free way. A delta update contains the nature of the update and is prepended to the rest of the state being updated. It enables making node changes without reading or writing the main part of stored data nodes. Particularly important, this avoids the need to read a page into main memory when performing blind updates, a feature common to RocksDB [13].

### 1.1 Record Caching

Cache hit ratio is critical to good performance when exploiting secondary storage for lower cost and durability, if it can be achieved cost effectively. Record caching does exactly that. Record caching would be difficult if delta updates were not supported. Cost effective record caching requires that we are able to avoid retaining in memory the secondary storage home pages for cached records

Based on our earlier cost-performance work, we show what the effect is of record caching. We do this by using a cost-performance graph[7] that clearly demonstrates how it provides lower cost along with better performance.

### 1.2 Concurrent Access

High concurrency when accessing system structures using multiple threads is critical to providing scalable performance under load. Latch-free technology is not commonly used but can avoid bottlenecking under highly concurrent access to hot data. Deuteronomy's initial techniques were far from optimal by failing to look past the basics of the technology for a more fundamental view of how latch-free techniques can and should be exploited.

We introduce a new methodology for using the latch-free approach. This approach focuses on minimizing duplicate work. It does this by pushing most new state construction until after conflicting accesses are resolved. The secret sauce is to make sure that conflict "loser" threads can do useful work while the conflict "winner" finishes constructing and installing the new state. In Deuteronomy, this is not possible without exploiting delta updates. We show how this approach works with our DC's Bw-tree[3] as an example.

## 2. Why RECORD Caching

The Deuteronomy TC is already very efficient, exploiting update log records both for multi-version concurrency control (MVCC) and as a record cache for updates. Records not currently in cache must be read from the DC. All cached records are stored in a record centric hash table. Reads that find their requested record in the TC hash-based cache have performance comparable to main memory databases.

Misses in the TC record cache are more expensive, sometimes dramatically so. If the record retrieval needs to access the DC's Bw-tree, then such a record retrieval has performance close to the performance of a conventional storage engine (key value store), which is dramatically higher. So it is no surprise that improving the TC record cache hit ratio can improve performance dramatically.

### 2.1 Caching Cost/Performance

Expanding the size of the TC record cache increases performance by increasing the cache hit ratio. Balancing cache execution cost versus the cost of its storage is an important aspect of providing not just a high-performance system but a cost-effective system as well.

We want to choose where we place data depending on how frequently we execute on it, when it can be either on SSD or in DRAM. We simplify the treatment here. A more complete treatment is in [7]. The costs that we include are the (rental) cost per unit time where all parts have the same lifetime. It includes rental costs for both storage and execution. Storage rental cost for DRAM is $M, for SSD it is $FL. Rental cost for the processor execution is $P, and for the SSD it is $I, and includes the processor I/O execution cost plus the SSD cost to execute an I/O operation. Hence $I = $P + $IO. A fuller discussion of this, including our estimates of these costs, is given in [7].

The storage costs do not change. The execution costs change with the rate ROP at which we execute operations on data. We compute the cost per operation when data is stored on SSD ($SS) and when it is stored in main memory ($MM). The point at which we want to move data from secondary storage to DRAM cache is when these costs are equal. The result is that $T_i$ (the time interval between data accessed) is inversely proportional to the size of the data unit being cached. The computation in [7] demonstrated that for a 4K size unit of storage (e.g. a data page), the break-even point when operation on cached data are the same cost as operations on flash data is around 45 seconds. We calculated that using formula (6) in [7], where $P_s$ is the unit size of the cached data.

It is important to emphasize that 45 seconds is the access interval for 4k bytes of data. If we cache records instead of pages, the interval becomes longer, inversely proportion to the size of the data unit cached, as the storage costs are lower. If there are 10 records per page, the record cache persistence is a factor of 10 longer, or 450 seconds (7.5 minutes) than for the page.

We illustrate both page and record caching in Figure X, using 2 records/page so that one can see the impact of the record storage footprint vs the page storage footprint. Using the smaller record storage unit extends the time in which it is cost-effective to keep records in main memory. The longer the time, the higher the cache hit ratio. The larger cache size enables performance closer to that of a main memory storage engine with lower miss ratios.

### 2.2 Managing Cached Data

A record cache not only can increase cache hit ratio by extending the time in which a record can remain in main memory. It can also improve the record access performance. In [5], we used hashing to access cached records. Hashing is many times (perhaps a factor of 100) faster than accessing a record in a page cache via a B-tree index. Using a B-tree like index requires access proportional to O(log(N)) where N is usually not the number of cached records, but the entire number of records in the file. Hashing has performance of O(1), with many fewer processor cache misses for discrete record requests.

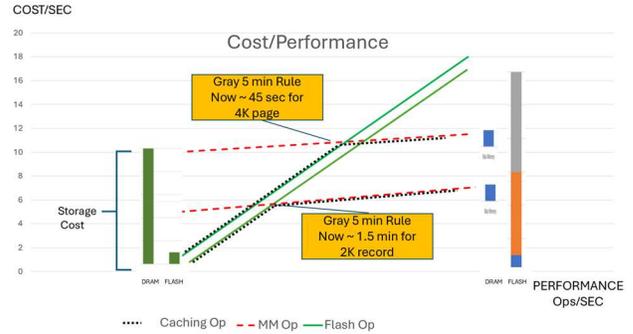

**Figure 1: Cost/Performance when one changes from 4K page caching to 2K record caching.**

Why then do database systems typically cache pages. There are, I believe, three reasons. (1) The number of I/O accesses supported by hard disks is extremely limited and a page is the unit of transfer in any case. (2) A page might yield a couple of access hits for this single disk I/O. (3) Database systems typically support range queries, which are surely not directly supported by a hash accessed cache. We return to this later.

The Deuteronomy record cache is in our TC. The cache contains both recently updated records, which begin records on the transaction log, and records read from the DC's Bw-tree. We use a common hash mechanism, managed in a log structured way, that can drop cold records, while continuing to provide hashed access to warm records. Since records may move in the log structuring process, we use a level of indirection that insulates our code from these movements. This is illustrated in Figure 2, taken from [5].

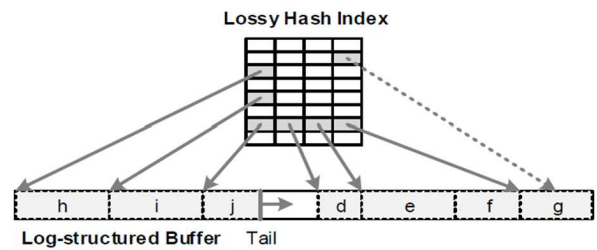

**Figure 2: The read cache is structured as two latch-free structures. A lossy hash index maps opaque 64-bit identifiers to offsets within a large log-structured buffer.**

### 2.3 List Handling

Deuteronomy performs a lot of list processing, involving pointers linking records together. For example (there are others), a hashing structure frequently includes linked lists of records "hanging off of" a hash indexed vector. Using log records in the record cache, together with records read from our Data Component, provides opportunities to share the record in multiple places via pointer connected list elements.  If one can avoid some of this pointer management, that is always a plus.  In a latch-free system, data management requires additional effort to provide pointer safety. Deuteronomy uses an epoch mechanism for this.

Not only is storage management expensive, but traversing linked lists can involve first accessing the pointer element, and then accessing the data associated with the pointer element. The pointers are managed separately, both in their allocation and in their garbage collection. This is a prescription for "slow". But it does not have to be that way.

What we suggest here is to prepare storage that is large enough to hold both list pointers and the associated list element.  This is by no means a new idea.  But we approached it in a rather round-about way.  Updates become records on the log.  There are pointers on the log, but they are, as is typical, pointers connecting updates in a given transaction.  However, we can allocate storage for additional pointers in the log.  These pointers will share a common lifetime with the data record adjacent to them.  And these pointers are particularly easy to garbage collect as the normal database log checkpointing can be used without further ado to reclaim storage.

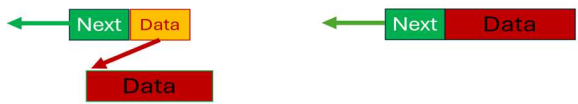

**Figure 3: List formatting of two forms. The left form requires 2X the allocations of the right and is more likely to trigger a processor cache fault when "Data" is accessed.**

With record reads from the data component, we can immediately move them to storage that is sufficient to hold not only the records but the pointers we expect them to need (e.g. hash list pointers). Again, these pointers share the same lifetime as their adjacent data element (record) and can by garbage collected with it, thus reducing the number of allocations that need to be managed separately.  This list management strategy can also be used in the data component when prepending updates to a list of prior updates.

## 3. Latch-free Concurrency

### 3.1 Efficiency Considerations

The Bw-tree is known not only for having good performance but for the use of latch-free technology which reduces (potentially) the execution cost when multiple threads are accessing, reading, and updating data on an index sequential access method.  I write "potentially" because there are a host of techniques that might be used for the specifics.

### 3.2 Latches

Using latches tends to make correct multi-threaded programming easier, by excluding multi-threading in the difficult sections where race conditions occur.  For example, if a thread is updating a B-tree node, setting a latch (a Test and Set) will ensure that no other accessor, reader or writer) will see the node in a transition state. Unfortunately, with the increased number of threads, this latch based critical section approach interferes increasingly with efficient use of the threads.  Basically, the threads need to either idle (usually a spin wait) for the latch to be released, or switch to another task that is able to be executed. But switching threads is very expensive, both in number of instructions and the impact on processor cache hit ratios. Both strategies are used in varying combinations. All result in interfering with the maximum exploitation of thread computational efficiency.  And in high contention work loads, latching seriously limits multi-thread scalability.

### 3.3 Latch-free Basics

Latch-free technology, usually using a Compare-and-Swap (CAS) instruction frequently avoids the kind of thread interference faced by latches and has the potential to more fully exploit thread processing capability.  Because access to data is not impeded, great care must be taken when state changes are made.  Essentially, a new state is created off-line and then installed by switching a pointer from old to new state atomically using a CAS. At least for a while, both new and old state need to be accessible, with old state discarded only after one can be sure that there is no further access to it by any active thread.  This requires the construction of an infrastructure to deal with this, e.g. an epoch mechanism that tracks whether an active thread continues to execute on some part of an older state.  This is non-trivial to implement correctly with high performance.  But once done [3], it plays only a very minor role in the cost of thread execution.

The Bw-tree node update illustrates how simple and efficient the latch-free approach can be, while also revealing that there are complications.  All nodes are accessed via a mapping table, the table index identifying the node and its entry pointing to the node state. References to nodes are always via the mapping table.

To update a node requires atomically replacing old node with new node.  This can be expensive.  But for a simple update, this is trivial. We update a node by constructing an update "delta" that points to the old state.  When installed, the update delta logically changes the node state to include the impact of the delta.  This is accomplished by prepending the update delta to point to the prior state).  This update is then installed using a CAS to update the mapping table so that the node's entry there includes now the new update delta.  Thus, updates are trivial, and the state change is effective immediately. Rarely, a competing update on another thread will race to install its update first.  The CAS used to install the new state arbitrates which thread wins the race.  The loser thread then repeats the process, now using the updated state that resulted from the winning thread's effort. The great advantage of this approach is that the new state shares the old state and minimizes the execution load for the updating thread.

### 3.4 Larger Node State Changes

There is an added cost for threads accessing a delta updated node, however.  Cost includes searching a delta list in addition to whatever read-optimized base state we started with. This is increasingly less efficient as the delta list grows.

So periodically, we want to incorporate the list of delta updates into the node base state. Our original effort worked as follows. To the side, we built a new state that consolidated the base state and the effects of the delta updates. Then, with a CAS, we installed a pointer to this consolidated state. The logic here is simple but there is a performance penalty. If multiple threads try to install a consolidated state, only one will win the CAS, which is what we want. But all the threads will pay the cost of consolidation, which is nontrivial. We want to avoid this extra cost. We introduce here an idea first introduced by Rui Wang [10]. The idea is for threads to race to install a delta instead of racing to install the consolidated state. This means that the losing threads only lose the work of generating and trying to install the delta.

The delta does not update a node's data state. Rather it announces that a thread is consolidating deltas with base state. We will use notices in other Bw-tree format changes as well. A thread consolidating deltas with base state first prepends a cNOTICE to the state of the node. The cNOTICE announces that the thread has signed up for the task of consolidating the node.

Constructing and prepending a cNOTICE is trivial compared with generating the consolidated state. And once posted, it prevents other threads from paying the cost of generating a consolidated state. Only the winner of the race to post the cNOTICE pays the cost of consolidation. Losing threads pay only the cost of posting the cNOTICE and failing. Since these are usually threads performing updates, they can continue execution "above" the cNOTICE to prepend their updates to the node state. Only the "winner" of the cNOTICE race pays the cost of node consolidation, on a state that is guaranteed not to be changed by any other thread.

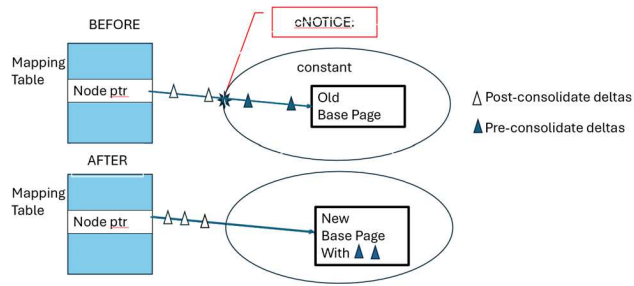

**Figure 4: Using a cNOTICE to consolidate a page. The thread that installs the cNOTICE is the only consolidator, while others continue to read and update the node.**

To ensure that the winner gets his update posted prior to updates of losers, the winner may build a two delta list to prepend to the node state, its update below the cNOTICE. This permits the cNOTICE to guard node state that includes the winner's update.

**3.5 Structure Modification Operations**

Larger multi-node changes require more cost prior to the CAS. These multi-node changes are structure modification operations (SMOs)[1]. Our goal remains to minimize pre-posting work of all competing threads. But, we want loser thread execution to continue smoothly. As with the cNOTICE, there is part of the prior state that can only be changed by the CAS winner. We want subsequent delta updates to act as if the SMO is completed. Below the notice, the old state awaits one or more additional transformations as specified by the SMO. As with consolidation, the expensive part of the SMO is replacing the state guarded by the notice.

**3.5.1 Node Split**

Node splits in the original Bw-tree were done in the same spirit as our prior consolidations. We created a new node for half of the old node entries, copied these entries from old node to new node, and then attempted to install a split delta using a CAS, with the usual B-link tree fix-ups to follow. Multiple threads may attempt this, meaning CAS losers have also performed this node building work.

We reduce the work done by a thread prior to the CAS by exploiting a split notice (sNOTICE). The CAS winning thread does most of the node split work. But there is some work that needs to be performed to ensure that CAS losers can proceed. We want to minimize that work while ensuring that state prior to the sNOTICE is protected from changes other than by the CAS winner. This is illustrated in Figure 5. We reduce this work and enable loser threads to proceed even before a half-split is completed. We do not include the parent index node update.

1. For new node N, we only make a slot in the mapping table. We allocate space in the I/O buffer for the data states for both N and O. But no data is moved. Allocating this space in the I/O buffer ensures that the split either persists completely or not at all should the system crash. The buffer is not written to storage until the splitter releases the buffer. Further, O's pre-split state, now split with N, will be in the log structured store buffer prior to further delta updates for O or N.

2. The sNOTICE contains the key we used to define the split and indexes to both N and O mapping table entries, and the location in the I/O buffer of the space allocated for O and N. The sNOTICE is posted without contention at N's mapping table entry. The split defining key becomes the new side link key for O.

3. We use a CAS to install an sNOTICE at O. This state is shown in Figure 5. If the CAS fails, we only lose the work in 1. and 2. above. No data has been moved, which is the major split execution cost. The sNOTICE is now in place at both O and N, permitting O's data below the sNOTICE to be shared by both. Subsequent accesses to O will encounter the sNOTICE, which directs them to O or N as indicated by the sNOTICE's split key. Accessors to either N or O execute "normally", except for sharing the state protected by the sNOTICE until the N is fully instantiated and O is cleared of entries that now reside in N.

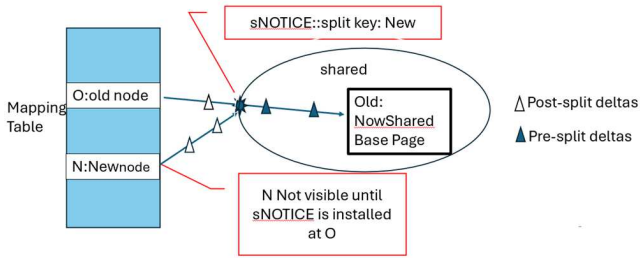

**Figure 5: Node split at step 2 – simplified. The encircled part is guarded by sNOTICE and is shared and read-only. Updates can continue to both old and new nodes.**

4. The sNOTICE CAS winner moves N's data (with keys larger than the sNOTICE split key) from O to N's allocated buffer space. Then the prepended deltas at N are connected to N's state in the buffer, ending the sharing arrangement.

5. The records in the shared state less than the split key are now moved to their new space in the buffer. We now replace the sNOTICE in O with a pointer to the newly formed version of O. We have now completed what has been called a half split, i.e. splitting the data between two nodes.

6. The final step is to update the parent index node with an index term that directs the search from the parent node to the appropriate node below it based on the split key. This is the second step in a B-link tree split. If the system crashes prior to accomplishing this, this part of the split can be completed during or after recovery.

### 3.5.2 Node Merge

Our original Bw-tree paper did not correctly describe the merging of nodes. The presence of notices makes the merging process easier. And it is even easier here when we restrict slightly the scope of the nodes that can be merged. We merge only those nodes that are not referenced via the lowest value index term of a parent index node. Merging data nodes eliminates an index term in the parent index node. We do not preclude parent index nodes from splitting. This creates a problem should the split process select to split the parent index node in the region where a node merge is occuring. When that occurs, both resultant index nodes of the split might be responsible for accessing part of the key space of the data nodes being merged.

With the introduction of a merge notice mNOTICE, this problem is avoided. The mNOTICE is posted to the parent index node P of the data nodes to be merged. It indicates which key space adjacent index terms reference the data nodes that are to be merged, hence identifying also the node to be removed (the higher order one D). It is posted using a CAS to ensure that P has not been changed.

Now the parent no longer logically contains an index term for D, and all searches for its data proceed via the lower order node L. L contains a side pointer to D and a search will follow the side pointer should the data in D be read or updated. This is the opposite of the node split case. During a split, one node contains data intended for two, while here two nodes contain data intended for one.

Other threads may be slow but still in the process of updating D. For this reason, we install an dNOTICE on D using a CAS. The dNOTICE directs these updaters back to L, where they can post a delta update on L. Finally, we consolidate the two pages into one by posting an xNOTICE on L. This consolidation brings both nodes together with all data below it being used, including the data in D. The xNOTICE protects the data below it from being changed as L and D are merged into a new L node.

Thus, we see that to isolate the parts of L and D that are being merged into a new L, we need two notices to ensure that the parts being merged are not changed during the SMO. And we need the third notice to ensure that the parent index node P does not split in an inappropriate way while the merge is going on. These notices are posted in order (1) mNOTICE to protect P's index space from being split between L and D, (2) dNOTICE to prevent direct access to part of D's data, (3) xNOTICE to prevent access to D via L's side pointer. Concurrent reads remain, and concurrent updates are all delta updates on L's state. L logically covers their combined part of the key space. And delta updates logically update the merged node, without interfering with data consolidation.

The heavy-duty part of the node merge is consolidating entries from L and D into a single optimized and contiguous storage area. And, as before, this is done after placing the required notices to protect the state from changing during consolidation. Losers of the race to post the mNOTICE can proceed with their reads or updates, assured that the winning mNOTICE thread will execute the node merge.

### 3.6 Thread Failures

The "all-at-once" state replacement at the time of the CAS, while more expensive in execution, does have the advantage that should the CAS winner thread die for some reason, there is always another thread that will do the work instead. However, we can achieve this result using notices as well. Instead of a notice identifying a permanent winning thread, the notice can have a time-out, such that after an interval, the notice will no longer stop another thread from doing the job previously awarded to the prior notice posting winner. One might use a clock time for this purpose, but using epoch numbers might be less expensive. A notice posting thread includes in the notice the epoch in which it is executing. After say two or three epochs, if the work has not been done, another thread can perform it instead.

### 3.7 The Notice Paradigm

There is a common thread (excuse the pun) through how we use notices to provide non-blocking concurrent read and write access for shared data under contention.

1. Supporting delta updates is essential so that an unchanging pre-delta state can be safely read, and the state can be updated while reads are active.

2. Notices are used to facilitate light weight conflict resolution when more than one thread wishes to make an update that is transformative as opposed to delta based. A notice is prepended in the same way as a delta update.

3. Notices are used to guard part of the state whose transformation (update) cannot be accomplished via delta updates. The state to which the notice is prepended is guaranteed to be unaltered by any other thread approaching that state along the delta list on which the notice is placed.

4. Notices must be prepended on all paths to the part of the state being transformed. This ensures that the notice protected part of the state cannot be changed except by the notice poster.

5. The state being transformed is read-only. It is transformed by replacement in an atomic action attaching it to surrounding state on all notice protected paths.

6. Latch-free infrastructure needs to be present to perform garbage collection when it is safe to do so. Deuteronomy uses an epoch approach [3].

As with other conflict resolution methods, e.g. latches, care must be taken. For example, with latches, deadlocks are possible without proper ordering among the latches. Care must also be taken with notices, their placement and their ordering, to ensure correctness.

## 4. Other Work

There are several other storage engines currently competing for attention. We used Berkley DB, a caching-based storage engine, in early benchmarking and its performance was not close to the Bw-tree [12]. RocksDB [13] is probably the most widely used storage engine, in part because it is open source. Also, its record updates can be executed without reading data from secondary storage, a valuable property shared by Deuteronomy. In our benchmarking, now several years ago, Deuteronomy performed better. But Deuteronomy is not open source, so it is used in only a few places [4,8,10].

One paper insisted on comparing the Bw-tree to purely in-memory access methods and, not surprisingly, found the in-memory methods faster [11]. Other main memory methods also demonstrated greater performance, but these papers did not assess cost-performance. We did a cost-performance comparison with Silo [9], a very fast main memory system, and found it to be faster, but it used much more main memory, leading us to the conclusion that only if you required stellar performance irrespective of cost would you choose it over Deuteronomy, which can move data from its main memory cache to secondary storage to reduce cost.

It is usually the case that hot data, which should be in the main memory cache, is substantially smaller than cold data, which should be on secondary storage until it is accessed and becomes hot. This is the way that caching-based systems balance cost vs performance, and why they are so widely used. All the large commercial database systems are caching based systems for exactly this reason. Customers want to reduce "COGS"—the cost of goods and services.

## 5. Conclusions

We describe two technologies. (1) Hashing based record caching gives near main memory system performance for individual record read and update operations in a cost-effective way within a caching-based system. Cache misses will, of course, occasionally add substantial cost to accesses. (2) Latch-free techniques are improved by our notice mechanism, to both avoid redundant work and to make certain large state changes such as occur in B-tree type access methods simpler and more efficient, avoiding most redundant work, and never requiring a thread stall or switch.

Short code paths, improving processor cache locality, and reducing data cache misses all work to improve system performance. But reducing thread switching is one of the surest ways to improve performance. A thread switch "trashes" the processor cache and can result in very substantial and often unseen overhead should the switching involve the operating system. Using latch-free techniques do not always have the shortest code paths when contention is ignored. But by providing a path for contention losers to continue with productive execution avoids the relatively huge cost of thread switching.

## Acknowledgements

An effort such as Deuteronomy features the work of many people over several years. Mohamed Mokbel, Alan Fekete, Justin Levandoski, Sudipta Sengupta, Ryan Stutsman, Rui Wang, and Kevin Zhao, and Umar Minhas all participated in the effort. Special thanks to Rui Wang for showing that node consolidation could be more efficiently done with a "notice-like" approach.